# Validation of Quantum Computing for Transition Metal Oxide-based Automotive Catalysis

**Yuntao Gu**
*Global Research and Development*
*General Motors Company*
Warren, MI, United States
yutnao.gu@gm.com

**Louis Hector, Jr**
*Global Research and Development*
*General Motors Company*
Warren, MI, United States
louis.hector@gm.com

**Paolo Giusto**
*Global Research and Development*
*General Motors Company*
Mountain View, CA, United States
paolo.giusto@gm.com

**Matthew Titsworth**
*Software and Services*
*General Motors Company*
Austin, TX, United States
matthew.titsworth@gm.com

**Alok Warey**
*Global Research and Development*
*General Motors Company*
Warren, MI, United States
alok.warey@gm.com

**Dnyanesh Rajpathak**
*Software and Services*
*General Motors Company*
Warren, MI, United States
dnyanesh.rajpathak@gm.com

**Eser Atesoglu**
*Research and Development IT*
*General Motors Company*
Warren, MI, United States
eser.atesoglu@gm.com

*Abstract* —Quantum computing presents a promising alternative to classical computational methods for modeling strongly correlated materials with partially filled d orbitals. In this study, we perform a comprehensive quantum resource estimation using quantum phase estimation (QPE) and qubitization techniques for transition metal oxide molecules and a Pd zeolite catalyst fragment. Using the binary oxide molecules TiO, MnO, and FeO, we validate our active space selection and benchmarking methodology, employing classical multireference methods such as complete active space self-consistent field (CASSCF) and N-electron valence state perturbation theory (NEVPT2). We then apply these methods to estimate the quantum resources required for a full-scale quantum simulation of a $Z_2Pd$ ($Z=Al_2Si_{22}O_{48}$) fragment taken from the $Pd/2(Al_xSi_{(1-x)})$ catalyst family where $x=Si/Al$. Our analysis demonstrates that for large Pd zeolite systems, simulations achieving chemical accuracy would require $\sim 10^6$-$10^7$ physical qubits, and range that is consistent with the projected capabilities of future fault-tolerant quantum devices. We further explore the impact of active space size, basis set quality, and phase estimation error on the required qubit and gate counts. These findings provide a roadmap for near-term and future quantum simulations of industrially relevant catalytic materials, offering insights into the feasibility and scaling of quantum chemistry applications in materials science.

## I. INTRODUCTION

Quantum computing offers an efficient approach for simulating materials with strong electron correlation, which are otherwise computationally challenging using classical methods [1–12]. In particular, quantum phase estimation (QPE) enables simulation of the full Born-Oppenheimer (BO) nonrelativistic electronic Hamiltonian with polynomial scaling, whereas classical approaches require exponential scaling. Battery materials, particularly transition metal-based solid-state systems, represent a major area of interest for quantum simulations [13–15]. However, simulating these materials on current quantum computers to the required accuracy is current infeasible. Rapid advancements in plane wave algorithms and quantum embedding techniques [11, 16–18] may improve feasibility, but currently, smaller molecular systems simulated with Gaussian basis sets provide a more practical path toward industrial relevance. Quantum resource estimates assume logical qubit error rates achievable in the next five years, providing a roadmap for practical quantum simulations of catalytic materials.

In this paper, we focus first on three binary transition metal oxide molecules, viz.; TiO (Titanium (II) Oxide), MnO (Manganese (II) Oxide), and FeO (Iron (II) Oxide), We then turn to a Pd zeolite catalyst fragment $Z_2Pd$ ($Z=Al_2Si_{22}O_{48}$), extracted from the $Pd/2(Al_xSi_{(1-x)})$ catalyst family where x is the Si/Al ratio (x <1). Each of these materials is of relevance to the automotive industry with the binary transition metal oxide molecules associated with catalysis, pigments, coatings, and strengthening of body structure materials. The catalyst is of interest for emissions reduction. These materials pose challenges for both classical and quantum chemistry simulations [19]. While this study centers on catalytic applications, our methods extend to broader areas, including battery materials modeling [15, 20]. We organize this paper as follows. First, Sec. II reviews classical catalysis modeling techniques. Then, Sec. III outlines our quantum and classical simulation approaches, followed by Sec. IV which validates our methods on simple layered transition metal oxides. Section V estimates quantum resources for the Pd zeolite fragment. Sec. VI summarizes the results.

## II. MODELING CATALYSIS

Catalysis design is largely experimental, relying on trial and error due to a vast chemical space. Computational modeling aims to provide predictive insights into catalyst properties such as activity, selectivity, and stability [22]. Electronic structure calculations help derive descriptors and kinetic rate equations, improving catalyst structure and reaction mechanism understanding. Transition metal catalysts are particularly challenging due to their multireference character, stemming from unpaired d electrons in frontier orbitals [14, 15, 20, 22, 26–30]. Density functional theory (DFT), though widely used, often fails to predict energy levels, oxidation states, and spin energetics correctly in these systems [15, 20–22, 26, 27, 29, 31,

32]. Two-state reactivity (TSR) is an important phenomenon in catalysis, where transitions between spin states influence reaction pathways [27]. For example, cytochrome P450 enzymes exhibit TSR, but their high-spin states make them difficult to model using DFT [33]. Studies have demonstrated that the complete active space self-consistent field (CASSCF) and density matrix renormalization group (DMRGR) approaches provide more reliable results relative to existing methods [34–35], but these methods are computationally expensive, and they scale poorly. Quantum computing provides an alternative approach. Quantum phase estimation enables rigorous ground-state energy estimation with polynomial scaling, making it attractive for catalysis modeling [2–12]. Variants of QPE can also be applied to excited and thermal state calculations.

### III.  METHODS

To determine our active space, we use the atomic valence active space (AVAS) [36] method. This enables automatic selection of molecular orbitals for multireference calculations by constructing orbitals capable of describing all relevant configurations of user-specified atomic orbitals. The atomic valence active space method has been validated for use with transition metal complexes [36]. We use the Restricted Open-Shell Hartree-Fock (ROHF) wavefunction orbitals as the starting point for our AVAS calculations. We further validated by comparing the AVAS CASSCF correlated energy to the correlated energy from active spaces selected using HOMO/LUMO and the unrestricted natural orbital (UNO) criteria for simple transition metal oxide molecules [37]. Each of these methods could be further combined with low-bond DMRG calculations of a larger active space to identify further molecular orbitals to include in the active space. For molecular geometries, we use literature values for the bond lengths of binary transition metal oxide molecules TiO, MnO, and FeO, and we use a DFT-computed optimal geometry for the Pd Zeolite fragment [19, 21].

For the CASSCF calculations, we employed Pyscf's FCI solver for the classical calculations. For our quantum computing calculations, we used the CASSCF method and all-electron (AE) benchmarking for the qubitization QPE simulation routines we propose [38]. Qubitization is a natural choice for simulating Pd Zeolite fragments on a quantum computer as it has favorable scaling and estimated runtimes for molecular simulations with Gaussian orbital basis sets. Furthermore, single, double, and tensor hypercontraction factorized Hamiltonians can be used with qubitization, drastically decreasing the computational cost of a simulation in trade-off for a controllable error penalty [39–41]. We note that Pd Zeolite fragments may also be amenable to plane wave simulation methods due to their quasi-periodic structure. However, we leave off benchmarking using plane wave methods as any simulation method using plane waves would require the use of pseudopotentials. A recent literature entry demonstrated that a QPE plane wave simulation would require judicious selection of functionals for electron-exchange correlation in periodic solids, but the source code used for resource estimation has not been open-sourced [11]. For all qubitization, Hamiltonian compression, and resource estimations used in this paper, we employed the OpenFermion package. We also used Pyscf to perform CCSD(T) calculations both within the active space and in the all-electron limit. CCSD(T) is not a multireference method and not typically used within an active space. We use CCSD(T) to estimate the error introduced by our Hamiltonian compression methods and to calculate the ground state energy for both the compressed and uncompressed molecular Hamiltonians. The energy difference between the Hamiltonians is taken as a more precise estimate of the error introduced via compression than the limit set by the norm of the Hamiltonian.

To recover the dynamic electron correlation energy, there are two general strategies: the diagonalize-then-perturb strategy and the perturb-then-diagonalize strategy. In the diagonalize-then-perturb strategy, one first attempts to capture static correlation for the zeroth-order Hamiltonian, and then proceeds to capture the dynamic correlation via a perturbative treatment of the remaining Hamiltonian. An example of this type of strategy would be a CASSCF + NEVPT2 (*N-electron valence state perturbation theory*) or CASSCF + CASPT2 calculation [42]. In the perturb-then-diagonalize strategy, the Hamiltonian is dressed first before diagonalization to account for interactions. We use Pyscf to perform our NEVPT2 calculations. CASSCF+NEVPT2 calculations are applied for the binary transition metal oxide molecules to demonstrate the efficacy of the method for recovering the correlation energy outside of the active space. While many applications discussed above do not require the full correlation energy, an accurate estimation as to which molecular geometries are most stable at different temperatures and pressures would require this. Furthermore, NEVPT2 relies on computing reduced density matrices (RDMs), which for sufficiently small RDMs can be readily computed on a fault-tolerant quantum computer (QC) (new NEVPT2 implementations like that used in Pyscf only require 3-RDMs) [43]. While this is possible on a QC, NEVPT2 (quantum or classical) requires computing a large number of reduced density matrices, a potentially costly proposition, especially on a QC. However, NEVPT2 is also the only fault-tolerant method for recovering the dynamic correlation energy of which we are aware of, but we anticipate that to change. Likely candidates for adaptation to QC would include range-separated methods for recovering dynamic correlation [44]. Range-separated DMRG+DFT is an example of the perturb-then-diagonalize approach. In range-separated DMRG + DFT, the Hamiltonian is separated into long and short-range correlation Coulombic forces. One uses the CASSCF method to compute the long-range dressed Hamiltonian energy and DFT to recover the short-range dynamic correlation. In all cases, the full correlation energy can be recovered with an all-electron calculation at increased computational cost. We employ the atomic valence active space (AVAS) method to construct efficient active spaces for transition metal catalysts [36]. Our classical validation involves CASSCF and NEVPT2 calculations, compared against CCSD(T) to quantify accuracy [37]. For quantum simulations, we use qubitization-based QPE, which reduces computational cost via Hamiltonian factorization techniques [38–41]. Dynamic correlation is addressed using two approaches: (1) Diagonalize-then-perturb, where static correlation is captured first (e.g., CASSCF + NEVPT2) [42], and (2) Perturb-then-diagonalize, where the Hamiltonian is modified before diagonalization (e.g., range-separated DMRG + DFT) [44]. NEVPT2 calculations require reduced density matrices (RDMs), which can be computed on a fault-tolerant quantum computer, but at high cost [43].

## IV. Modeling Simple Transition Metal Oxides

To benchmark feasibility, we validate these methods on simple transition metal oxides TiO, MnO, and FeO, before applying them to the selected Pd zeolite catalyst fragment. We choose these molecules since they are sufficiently small such that we can employ all classical simulation methods (other than for all-electron CASSCF calculations). These molecules require careful consideration of electron correlation; however, they appear to be predominantly dynamically correlated, as single-reference methods such as CCPlots and CCSD(T) describe them well. [21]. They allow us to validate the use of Gaussian basis sets. In Figure 1, we plot CASSCF + NEVPT2 energies for TiO using the ccpvtz basis set and three different methods for selecting the active space. For comparison, we also include the all-electron CCSD(T) and DFT with PBE functional energies. We note that for simple transition metal oxide molecules, CCSD(T) can be taken as accurate; CCSD(T) captures 99% of the correlation energy of transition metal oxide molecules [21]. The DFT energy is approximately one order of magnitude further from the CCSD(T) predicted correlation energy than all the CASSCF + NEVPT2 energies. For sufficiently large active spaces, CASSCF + NEVPT2 can be systematically improved.

Next, we compare the estimated quantum resources required to simulate the AVAS computed active spaces for the simple transition metal molecules. Comparisons are made under a matrix of varying conditions including molecule, active space size, norm of truncated active space Hamiltonian, basis set quality, and phase error for both the single and double factorized molecular Hamiltonian with qubitization QPE. We present our results in the plots below. We plot the results for the active space size heavily influences both the number of logical qubits required and the Toffoli gates while the desired phase space error largely only influences the Toffoli gate count. As observed in Ref. [10]. A higher basis set quality does not influence simulation cost due to a higher basis set quality, leading to a more complex Hamiltonian.

In Figure 2, we plot the results number of logical qubits (a) for the binary transition metal oxide molecules (by transition metals among the horizontal axes) and number of Toffoli gates (b). Figure 4 shows the variations of the number of logical qubits (a) and the number of Toffoli gates (b) for MnP with 1mHaphase error using a double-factorized Hamiltonian in the ccpvtz basis set. For each active space, we select the factorization of the Hamiltonian where the CCSD(T) error between the factorized and unfactorized Hamiltonian is closest to but still less than the maximum phase error. We note that, as expected, the active space size heavily influences both the number of logical qubits required and the number of Toffoli gates, while the desired phase error largely only influences the Toffoli gate count. As observed in Ref. [2], a higher basis set quality does influence the cost of simulation due to a higher basis set quality leading to a more complex Hamiltonian.

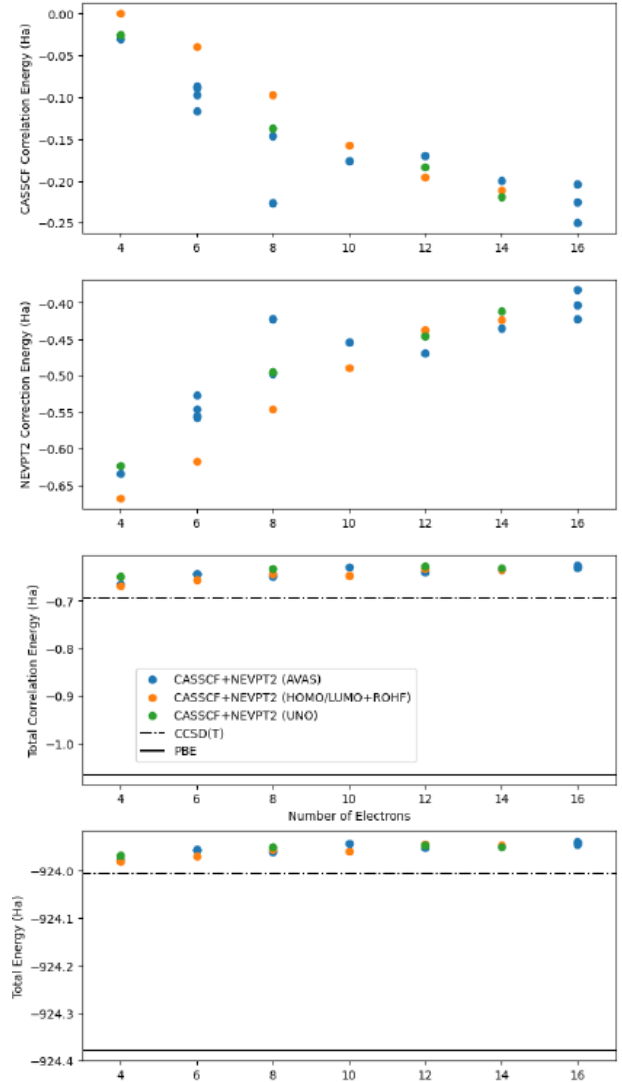

*Figure 1: Results for TiO: (a) top, active space correlation energies, (b) second from top, NEVPT2 dynamic correlation energy, (c) third from top, total correlation, energy, and (d) bottom, total energy using the ccpvtz basis set.*

## V. Modeling Palladium Zeolites

Having validated our methods for resource estimation on simple transition metal molecules, we move on to modeling the Pd zeolite catalyst fragment $Z_2Pd$ ($Z=Al_2Si_{22}O_{48}$), extracted from the $Pd/2(Al_xSi_{(1-x)})$. We initially began with larger zeolite fragments but were unable to converge our CCSD(T) calculations (hence the reason we are using a quantum algorithm for energy estimation). To estimate the energy error of our QPE calculations, we use the ERI Hamiltonian norm error discarded during the factorization procedure.

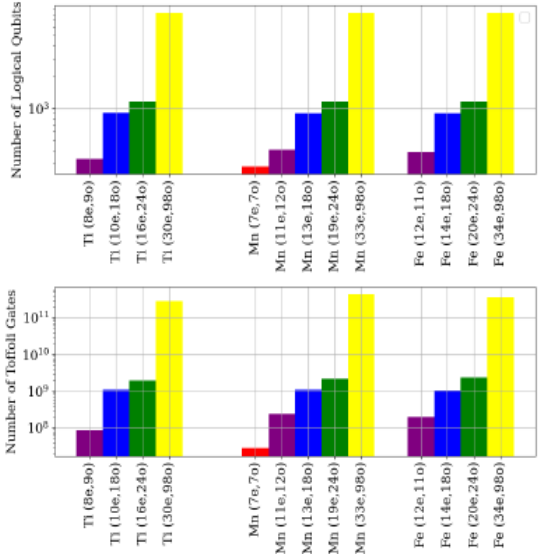

*Figure 2: For the binary transition metal oxides, plots for the number of (a) logical qubits and (b) Toffoli gates to perform qubitization QPE with 1 mHa phase error using a double-factorized Hamiltonian in the ccpvtz basis set.*

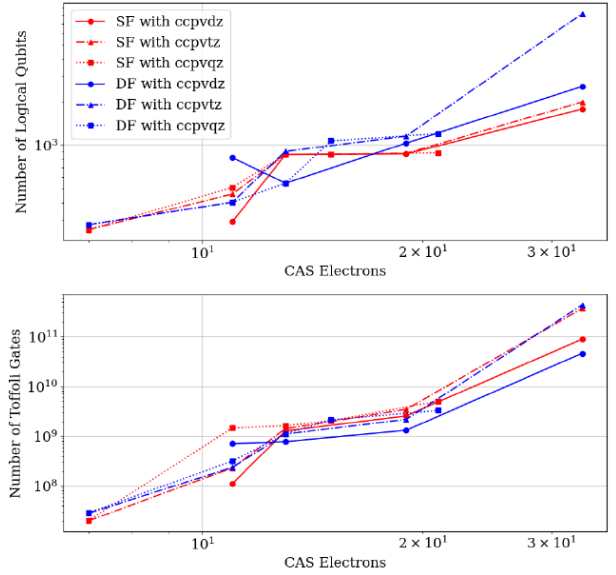

*Figure 4: For MnO, plots for the number of (a) logical qubits and (b) Toffoli gates to perform qubitization QPE with 1 mHa phase error using a double-factorized Hamiltonian in the ccpvtz basis set.*

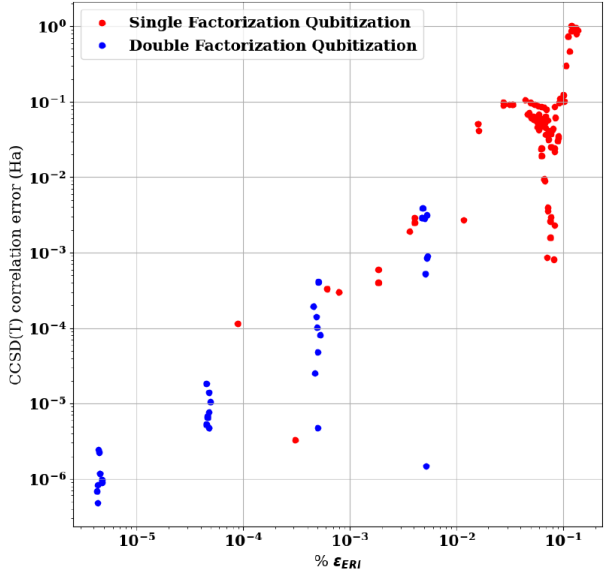

*Figure 3: Plot of the CCSF(T) correlation energy vs. %ERI for factorized Hamiltonians of the Pd zeolite catalyst fragment $Z_2Pd$ ($Z=Al_2Si_{22}O_{48}$).*

In Figure 3, we plot the CCSD(T) correlation energy error (with respect to the correlation energy of the non-factorized Hamiltonian) for the single and double Hamiltonian factorization methods using different factorization cutoffs vs. %ϵERI for the Pd zeolite fragment. The fragment is chosen with the Pd center having nearest and next-nearest neighboring atoms. The free bonds of the neighboring atoms are hydrolyzed. We observe strong correlation between the %ϵERI and the CCSD(T) correlation energy error for both methods. We chose %ϵERI < $10^{-3}$ as a sufficiently small %ϵERI to achieve a 1 mHa error in our QPE calculations for both single and double factorization methods.

Next, using the %ϵERI < $10^{-3}$ threshold, for a 1mHa error, we plot resource estimates for the Pd zeolite fragment ($Z_2Pd$ ($Z=Al_2Si_{22}O_{48}$),), the full Pd zeolite ($Pd/2(Al_xSi_{(1-x)})$), and the full Pd zeolite with 4-H2O molecules as an estimate for the cost of simulation. In Figure 5, we plot the quantum resources required to simulate the molecules to 1mHa accuracy. We plot estimates for several AVAS-selected active spaces (increasing the number of atomic orbitals used in active space selection) of the Pd zeolite fragment and of the full AVAS-selected active space (by including all potentially important atomic orbitals) of the catalyst structures. We plot both the logical and the qubit and Toffoli gate count for the full QPE circuit required for simulating the catalyst molecules to the desired precision using ccpvtz basis set with a circuit probability success rate of 90%. In addition, we plot the physical qubit count and the QPE circuit duration assuming a two-qubit gate stochastic error rate of $10^{-4}$ and a 10 μs gate time. We note that these are reasonable error rates, gate times, and QEC code architectures for near term superconducting qubit platforms.

In Figure 6, we plot the physical qubit count and circuit duration for simulation of the Pd zeolite fragment using two different qubit error rates. As expected, the increased QEC code distance required to sufficiently reduce the logical error rate largely affects the number of physical qubit counts needed but not the circuit duration.

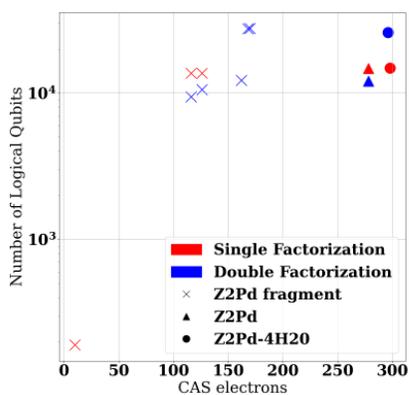
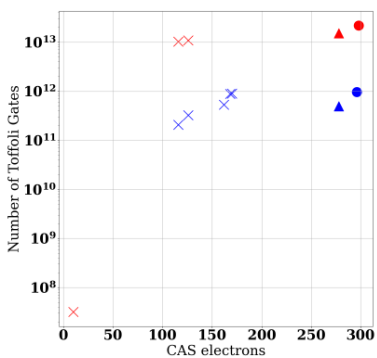
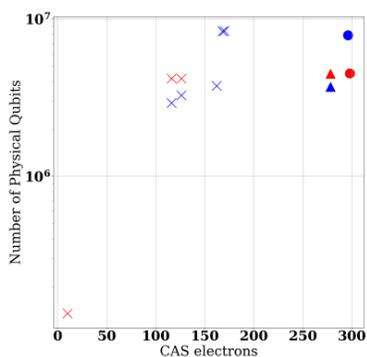
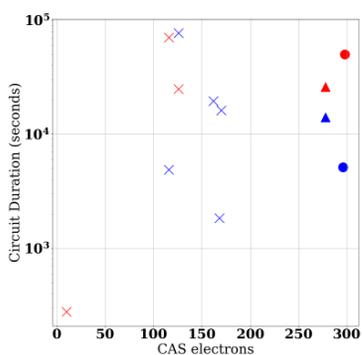

*Figure 5: Resource estimates for catalyst systems. (a) top, number of logical qubits, (b) second, number of Toffoli gates, (c) third, number of physical qubits, (d) fourth, circuit duration (seconds).*

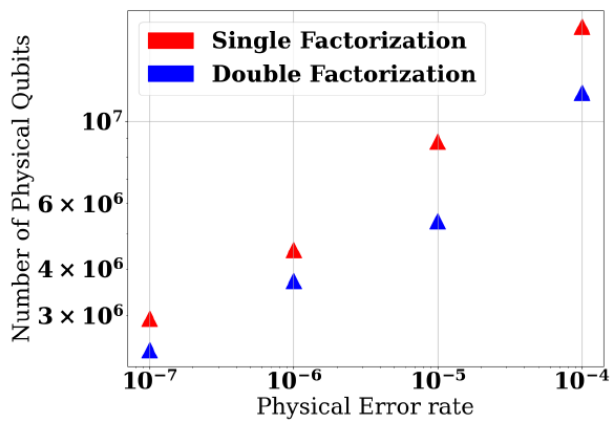
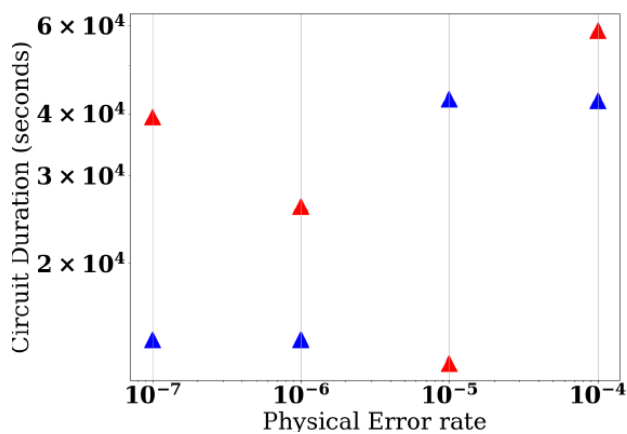

*Figure 6: For Pd zeolite fragment: (a) physical qubit count, (b) circuit duration assuming a two-qubit gate stochastic error rate of $10^{-4}$ and a 10 μs gate time.*

## VI. CONCLUSIONS

In this paper, we have presented an end-to-end analysis of the quantum resources required to accurately calculate Pd zeolite fragment ($Z_2$Pd (Z=$Al_2Si_{22}O_{48}$)) ground state energies to chemical accuracy. We note that while the specific zeolite chosen may not be of direct industrial interest, the broader problem studied—heterogeneous catalysis—has important applications in industry, ranging from battery chemistry to materials manufacturing. Our results provide a framework for evaluating potential applications in the future. First, we identified a gap in the industrial state-of-the-art for modeling chemical systems; the lack of sufficiently accurate simulations for large molecular systema is impeding progress in heterogeneous catalysis design. Next, we identified similar, smaller chemicals that allowed us to assess and validate our methods. Finally, we applied our methods for quantum resource estimation to the full zeolite problem. We found that even for the largest Pd zeolite unit cells, approximately $10^6$–$10^7$ physical qubits would be required at error rates that can be reasonably expected in the next five years (for architectures like superconducting qubits) or are already available (in the case of trapped ions). While these resources are several orders of magnitude larger than current QCs, a few QC hardware providers have these scales on their current technical roadmaps. We also expect that further theoretical advances in computational chemistry/materials science will further reduce the simulation requirements.